# Fainter and closer: finding planets by symmetry breaking

Erez N. Ribak[1,2,*] and Szymon Gladysz[2]

*1 Physics Department, Technion – Israel Institute of Technology, Haifa 32000, Israel*
*2 Applied Optics Group, School of Physics, National University of Ireland, Galway*
*\*Corresponding author: eribak@physics.technion.ac.il*

**Abstract**

Imaging of planets is very difficult, due to the glare from their nearby, much brighter suns. Static and slowly-evolving aberrations are the limiting factors, even after application of adaptive optics. The residual speckle pattern is highly symmetrical due to diffraction from the telescope's aperture. We suggest to break this symmetry and thus to locate planets hidden beneath it. An eccentric pupil mask is rotated to modulate the residual light pattern not removed by other means. This modulation is then exploited to reveal the planet's constant signal. In well-corrected ground-based observations we can reach planets six stellar magnitudes fainter than their sun, and only 2-3 times the diffraction limit from it. At ten times the diffraction limit, we detect planets 16 magnitudes fainter. The stellar background drops by five magnitudes.

## 1. The trouble with planets

Planets in other stellar systems (exoplanets) are difficult to discover, and even more difficult to image. This is because there is a very large intensity difference between them and their parent stars, which can be anything from $10^{-4}$ in the infrared down to $10^{-10}$ and fainter in the visible regime, depending on the physical properties of the system [1,2]. Light from the star is scattered off-axis by the atmosphere, the aperture of the telescope, the secondary mirror holders (spiders) and other optical elements in the path to the imaging camera. Techniques such as adaptive optics (AO) and coronagraphy aim to remove this effect by concentrating light back on the optical axis and removing it [3]. All future ground-based high-contrast imaging systems will employ coronagraphs and AO [4,5].

It has been shown through simulations and experiments that the limiting factors in high-contrast imaging from ground and space are static and quasi-static speckles [6-8]. While residual atmospheric aberrations after AO correction are random and will average out over time, these persistent speckles will stand out against the AO-corrected stellar halo and masquerade as faint sources even after long integrations. This is why future efforts to directly image planets will also entail post-processing schemes based on the concept of point spread function (PSF) subtraction. Proposed approaches utilize PSF estimates provided by on-sky rotation [9], as well as spectral [10,11] and polarization-based [12] discrimination between the light coming from the parent star and the companion. The problem with some of these techniques is that the PSF-subtracted images still contain static speckles (at a lower brightness level than in the direct images) due to errors in PSF estimation. These errors arise due to the inherent sensitivity of PSF subtraction to changes in "seeing", mechanical flexure or the introduction of extra imaging channels. Recently a detection algorithm has been proposed which relies only on multiple exposures and statistical properties of AO-corrected intensity [13].

In space missions efforts to calibrate out persistent speckles will concentrate on PSF estimation from wave front sensing and subtraction [8], as well as "speckle nulling" algorithms creating "dark holes" around the center of the image, thus improving the detectability of planets [14,15].

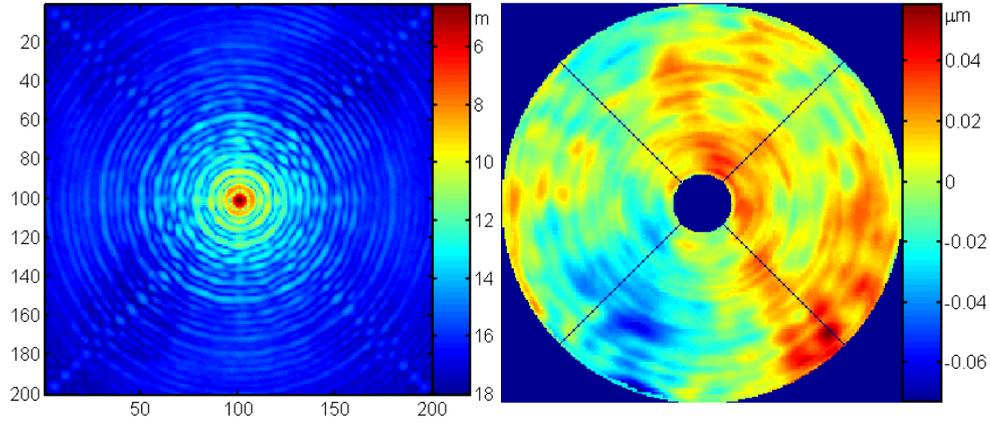

Fig.1. Simulated stellar diffraction pattern (left), including residual adaptive optics and polishing errors from a VLT 8 m mirror (right). This diffraction pattern is the sum of 200 exposures, each 0.1 s long, shown in stellar magnitudes (negative logarithmic scale, base 2.5) to reveal the faint off-axis structure. The Strehl ratio in this image is 85%. The scale on all images is λ/D = 4 pixels.

## 2. Symmetry and the image

When we examine a typical stellar image, it becomes clear that the light pattern from the star is not only dominant and difficult to reduce, but is rather symmetric (Fig. 1). This is because most telescope pupils have circular symmetry, with very few having hexagonal symmetry. Apart from spiders, almost all other parts of the system are rotationally symmetric, therefore the pattern of light on the detector maintains this symmetry, typically an Airy pattern ($\sim J_1^2(r)/r^2$). Deviations from symmetry only occur at a much lower level, after correction by AO and reduction by a coronagraph which might not be perfectly aligned or adjusted. Residual mirror polishing errors (which tend to be rotationally symmetric anyway) and aberrations downstream the optical system are modulated by the symmetry of the Airy pattern in the small-aberration regime. "Pinned speckles" [6, 16] form an additional rotationally symmetrical component in the focal plane because they are modulated by the Airy pattern. All of these effects contribute to confusion when trying to identify planets among many object-like features, having the same spatial scale ($\lambda/D$ - the wavelength divided by the aperture width) and shape.

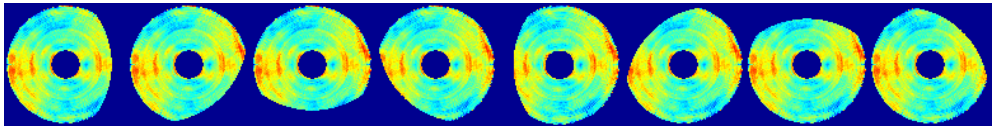

Fig. 2. Breaking the symmetry of a telescope by masking 20% of its pupil (both in area and diameter) at eight orientations. Notice the typical residual aberrations, on the scale of a few nanometers (SOAR telescope, courtesy A. Tokovinin).

We suggest here a simple method to overcome this problem, which breaks the symmetry of the optical system. We position a round masking aperture off-center and conjugate to the main telescope pupil, thus narrowing the pupil in one direction. As a result, the typical Airy pattern created by the star at the focus will be slightly wider and slightly shifted to the side. Rotation of this eccentric pupil mask about the telescope axis will rotate the distorted Airy pattern and move the speckles in the image plane. Since the images of the star and planet will not move, the location of the planet will be revealed at different mask orientations.

By how much do we want to reduce the pupil? Enough to cause speckle movement by full speckle size, while not reducing the resolution so as not to smear the stellar and planetary images. Since the speckle size is close to $\lambda/D$, and since the inner, brighter Airy rings of the stellar image are also separated by nearly that value, we need to cause a shift of about this distance. By limiting the aperture diameter A in one direction to $A \lesssim D$, the speckle pattern size will grow slightly to $\lambda/A$ and the shift will be of the same order. For example, to shift the second zero of the Airy pattern to fall on the third maximum requires $A/D \approx 2.22/2.65 \approx 0.84$. We found that indeed 20% pupil reduction is more than sufficient, although this value needs to be optimised for different planet locations and other coronagraphic diffraction patterns: the boiling of the speckles and the diffraction pattern increases with the distance from the center and with the amount of eccentric obscuration (Figs. 3, 4).

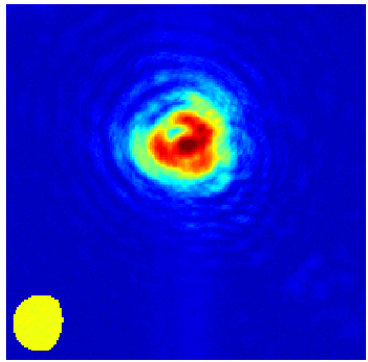

Fig. 3. Rotation of the diffraction pattern with the rotation of the aperture mask. A spatial light modulator imposed the masks as in Fig. 2 (inset), and we took the corresponding images of the laser diffraction pattern (log scale). The residual system and turbulence aberrations were slightly more than one wave length. (1.3MB Movie: http://physics.technion.ac.il/~eribak/LabPupilsDiff.avi )

Notice that we break the symmetry of the telescope and of the persistent speckle at the same time, while maintaining the rest of the optics intact. This is the reason that the stellar and planetary positions remain unchanged despite the broken symmetry, while their diffraction patterns are affected. We have experimented (in simulation) with different rotation angles, and with different image processing techniques. We found that four rotations by π/2 are sufficient for planets $\gtrsim 8\,\lambda/D$ away from the center, where the intensity symmetry is broken and mostly influenced by static speckle. However, for planets closer to the center and affected by the Airy pattern, finer rotations were required. Combined with other analysis techniques described below, these rotations allow getting as close as $2.5\lambda/D$ to the center. Notice that due to the fact that we measure the intensity of the light, not its amplitude, there are angular symmetries by the time the pupil rotates by π, which can further be used.

Fig. 4. Part of the Airy pattern: the first to third rings move as the width of the aperture changes

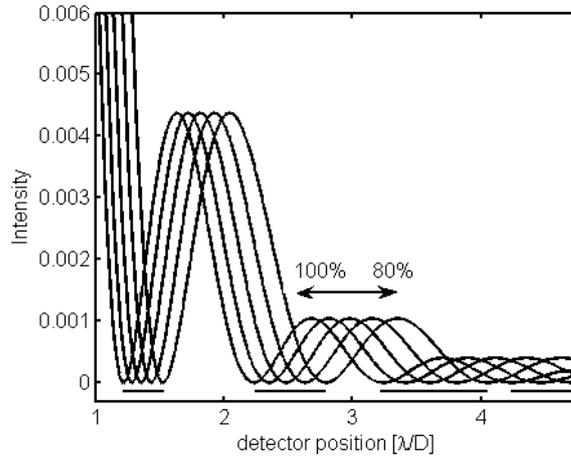

from 100% to 80% during the rotation of the masking pupil. The zero-intensity areas between the successive rings sweep larger and larger swaths (bars at bottom) allowing for faint planets to show up beneath them.

## 3. Simulations

We modelled SPHERE, the planet finder instrument planned for the Very Large Telescope [5, 17] using the PAOLA AO simulation package [18]. The approach to simulate AO long exposures, containing static speckles, is described in [13]. Briefly, PAOLA calculates the spatial power spectral density of the AO-corrected phase. We used this power spectrum to generate AO-corrected wave fronts. In [13] we modelled static aberrations by adding a mirror error map, such as in Fig. 1, to these residual wave fronts. Here we used the same approach to model the primary mirror, and on top of that we added a realization of an $f^{-2}$ spectrum to simulate aberrations from additional optical components [6]. The standard deviation of the primary aberrations was 20 nm, and 10 nm in the case of the $f^{-2}$ spectrum. The coronagraphic module of the planned system was not included in our model.

We briefly describe the differences between this work and the earlier simulations. The imaging wave length was set to 1.6 μm, because this wavelength is often used in spectral differential imaging based on the methane absorption feature of giant planets (we make no use of SDI here). The band width was 30 nm. The global tip and tilt were not filtered digitally out from the wave fronts. We generated 200 short ($dt$ = 0.1 s or 0.5 s) exposures from the wave fronts where the VLT pupil was masked by the rotating aperture (Figs. 1, 2). Between these exposures the eccentric aperture was rotated by 9°, totalling 5 cycles per data-cube. For comparison we simulated a direct long exposure, with no aperture obscuration (Fig. 1). Poisson, background (14 mag arcsec$^{-2}$) and readout noise (10 e$^-$) were added to all short and long exposure images in our simulations.

The modelled systems consisted of the primary star of magnitude 4, and 36 artificial faint companions, all having the same aberrated shape as the star, inserted in the field before adding the noise. It should be noted that the predicted dynamic range of SPHERE is larger than the values reported here due to the light suppression by the coronagraph which we did not model [17]. The focal-plane sampling was $0.25\lambda/D$, where $D$, the diameter of the VLT, is 8.2 m [19]. The planets were placed in a central cross, separated from the star and each other by 10 pixels, or $2.5\lambda/D$, or 0.1", and their intensity dropped by one stellar magnitude each from the center outwards, from 12 to 20 (each magnitude is 2.51 times fainter than the previous one). Notice that their location nearly coincides with the Airy rings. The brightest ones are barely visible in Fig. 1.

## 4. Distant planets

We first deal with the case where we wish to discern the persistent speckle from the planet when it is a many times λ/D from the star. It turns out that when we rotate the aperture, the persistent speckle undergoes a sinusoidal modulation to first order. All that we have to do is find the amplitude of the speckle intensity modulation. Once we remove that intensity, the planet will show up, as we show now.

The intensity at every image position x, y and time step t (with equivalent rotation angle $\omega t$) can be written in general as

$$I(x,y,t) = I_p(x,y) + I_s(x,y,t)$$
$$= I_p(x,y) + I_m(x,y)\{1 + \cos[\omega t + \varphi(x,y)]\} \quad (1)$$

where $I_p$ and $I_s$ are the planet and speckle (or stellar) contributions, $I_m$ is the amplitude of the speckle intensity variation, and $\varphi$ is an unknown phase angle. $I_p \neq 0$ only at the planet location. Similar to the fringe detection method [20] we collect four images, $I_a, I_b, I_c, I_d$, one at every quarter turn, namely when $\omega t = \pi/2, \pi, 3\pi/2, 2\pi$ (we drop the position dependence for simplicity). These four images are used to find the phase, since for every position $(x,y)$ we have $\tan\varphi = (I_c - I_a)/(I_d - I_b)$. If all the speckles had the same phase angle $\varphi$, we could have taken another frame where $\omega t + \varphi = \pi$ (Eq. 1). Thus we could have found the planet signal alone by rotating the mask to that orientation angle. But as at each position $(x,y)$ the phase is different (possibly even across the planet image), we need to take many more frames for all possible positions.

There is a much simpler solution, and that is to find also the amplitude of the speckle movement. Again from Eq. (1) we can see that regardless of the unknown phase, we get

$$I_m^2 = (I_a - I_c)^2 + (I_b - I_d)^2 - I_t; \quad I_t = I_a + I_b + I_c + I_d, \quad (2)$$

where the last term, $I_t$, is due to the skew of the governing Poisson statistics, frequently encountered in astronomical interferometry [21,22]. Finally we get $I_p = I - I_s = I_t - I_m$.

## 5. Close to the core

Things are not quite so simple when we get close to the center of the image, near the diffraction pattern, or what is left of it after nulling or scattering out. Here the movement of the pattern, as a function of orientation of the apodizing aperture, decreases as we get closer and closer (Fig. 4), and becomes more involved. We generalise Eq. (1) to

$$I(x,y,t) = I_p(x,y) + I_m(x,y)\{1 + P[\omega t + \varphi(x,y)]\}, \qquad (3)$$

where $P(x, y, t)$ is some periodic function of time, different from pixel to pixel. As an example, we show in Fig. 5 the behaviour of the intensity over three full revolutions of the aperture at two image pixels. While the movement is periodic, it can hardly be described by a simple sinusoid as in Eq. (1). Four measurements, or even eight (for two sinusoids), at different mask angles, might not suffice to describe it.

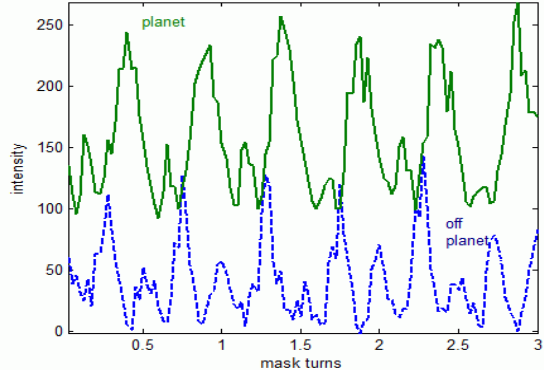

Fig. 5. Intensity in two pixels near the star during three pupil revolutions. One trace is elevated due to the presence of a relatively bright planet. The periodicity is not perfect due to residual tip-tilt errors, Poisson, sky, and read out noise.

Since the minima of stellar light contribute less background to the planet's signal, these are the best places to look for it, but the rotation angles, where these minima occur, are not well known. We resort to another strategy, which is to locate the minima for every pixel in this complex but repetitive pattern. It is quite clear that if we happen to take a measurement when the speckle signal $I_s$ is minimal, the total intensity $I$ will be minimal at the same time (Fig. 6). At the positions where the planet is absent the minimum during many revolutions should be close to zero or close to the persistent pattern, which doesn't move, while at the location of the planet the minimum will be close to the planet's intensity. Notice how similar this now becomes to Labeyrie's 'dark speckle' method [23,24], with the marked difference that we do not rely on phase changes to null occasionally the persistent speckle. Rather we seek these minima actively by modulating the aperture orientation and sampling the images frequently. Thus the algorithm simply forms an image of the minima in the focal plane over many revolutions. When we get close enough to the stellar pattern center, there might be no points where the speckle intensity drops to zero (Fig. 4). In these positions, the static speckles and diffraction pattern are still visible.

It should be mentioned that the dark speckle method suffers from an inherent bias. Even assuming that we choose exactly the right mask orientation without any stellar speckle present, we are still looking for the weakest signal. Rather than giving us the average flux from the planet, we get the minimum flux, which for Poisson statistics, and for read out noise of $\sigma$, is $\sim I_p - \sqrt{I_p} - \sigma$. The more observations we take, and the fainter the planet is, the worse the bias. Sky background will deteriorate this bias even further. A partial solution we adopted was to average the faintest occurrences in each location, rather than picking the absolute faintest one (Fig. 6). At any rate, this negative bias will only affect the photometry of the planet and might be accounted for statistically.

In order to avoid correction for the dark speckle it would be easier to characterise and remove the spectral behavior of $P(x, y, t)$ in Eq. (3). By Fourier transforming $I(x, y, t)$ only in the time dimension, we can locate the main temporal components which appear at known frequencies (e.g. 3, 6 and 12 cycles for the traces in Fig. 5). Then their amplitudes can be re-moved using Eq. (2). Initial efforts in this direction were partially successful because of residual tip and tilt errors. This Fourier notch method is somewhat similar to low-order polynomial fitting as proposed for wave length demodulation of star-planet systems [25,26].

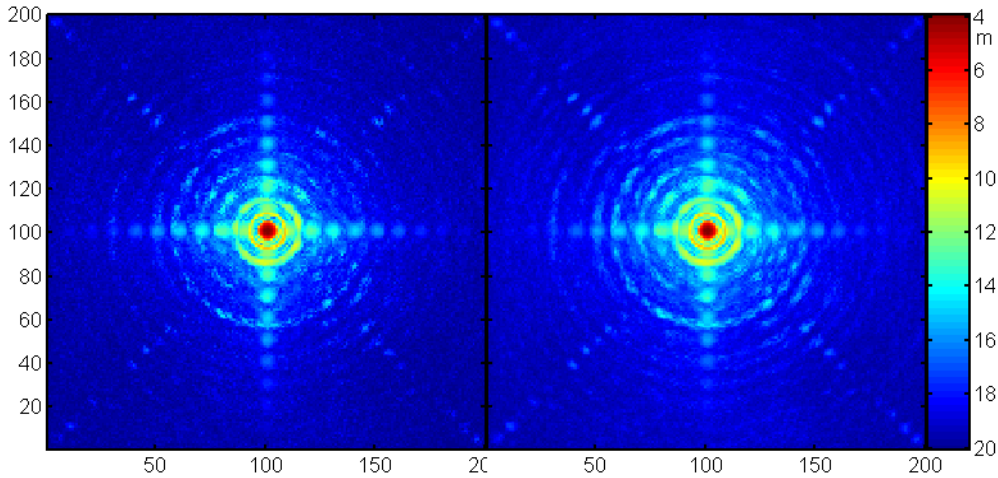

Fig. 6. The minima in every image position near a 4[th] magnitude star during five cycles of pupil rotation. Nine planets of magnitudes 12 to 20 were placed on four sides of the star (compare to Fig. 1). (Left) The absolute minimum in each image pixel; (Right) The average of the four faintest measurements. The image is smoother but contains more speckles.

Other methods can be employed for detection of the weak planet in the induced speckle motion, such as Wiener deconvolution by the known stellar PSF, PSF subtraction, stochastic speckle discrimination [13], and more. These methods can be performed in combination with other methods to further improve detection or reduce the chance for false alarm.

## 6. Further redundancies

We tried a few obscuration values for the masks of the shapes of Fig. 2. Each such obscuration percentage is more efficient for some speckle, while some other speckle or residual Airy pattern may prove more resilient. In order to counter that, we removed these persistent speckles by applying a combination of masks. For example, first we ran 200 rotation steps of $9°$ at an obscuration ratio of 11%, then 200 more at 16%, and so on. Even without changing the step size, this combination of the variable obscuration was very efficient in minimizing the stellar background (Figs. 7, 8).

Mechanically, changing the obscuration ratio cannot be performed any more with a simple stepper motor. Rather, it has to be done with by what is called a planetary gear, a name appropriate for this case. This gear can be non-circular or an axis-displacing one. Instead of changing the mask percentage after a number of revolutions, it changes the percentage at the same time as the rotation step, but completing a full cycle only after a much larger number of steps (Fig. 9). The corresponding frequency response will be more complex, but still describable with a manageable number of discrete frequencies.

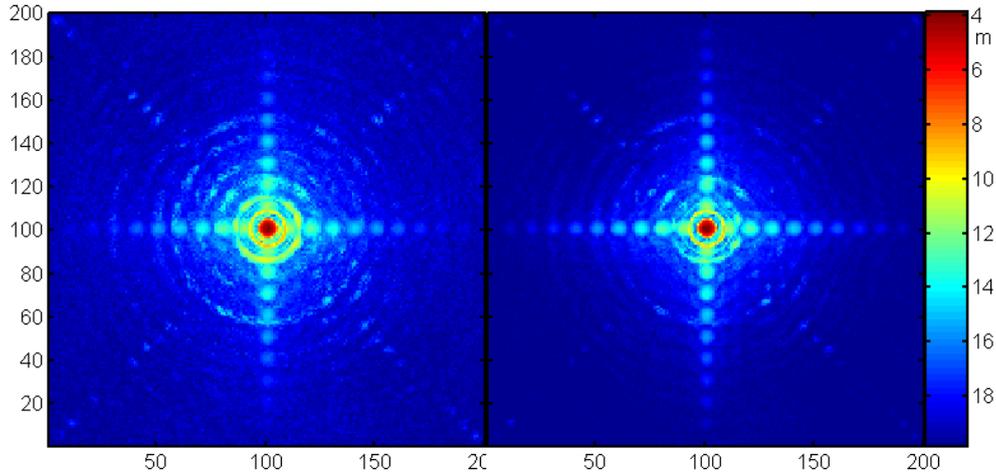

Fig. 7. (Left) The same image as in Fig. 6 Left. (Right) Many persistent speckles have now disappeared, by adding to the 31% diameter obscuration also 26%, 21%, 16%, and 11% obscurations, each 200 measurements lasting 0.1s every 9° rotation.

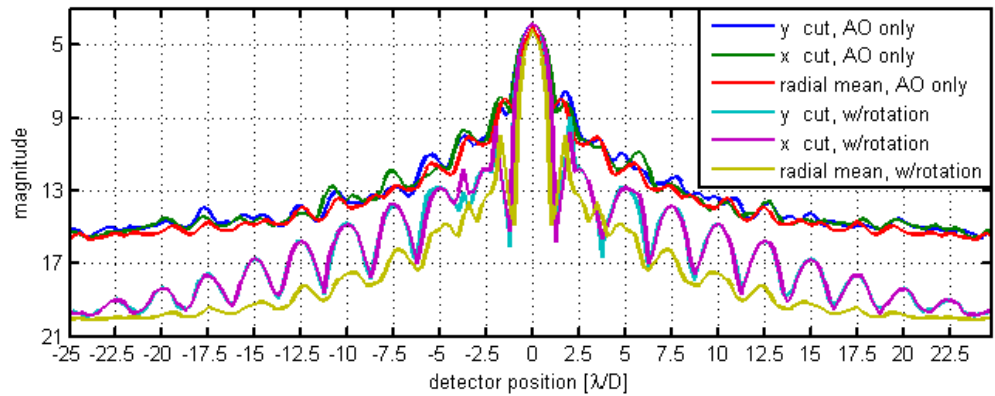

Fig. 8. For comparison, we take the long-exposure adaptively-corrected pattern (Fig. 1, Left), and calculate the radial mean (average over angles) and cross cuts through the Airy rings (top three curves). Then we do the same for the pattern of minima after aperture rotations at different obscurations (Fig. 7, Right). The residual Airy rings have nearly disappeared (bottom three curves). The horizontal and vertical cuts are very similar (middle two curves), and stand out against the radial mean (bottom curve). The planets are clearly visible even though we set them almost on top of the Airy rings, every 2.5$\lambda/D$. The difference between radial means shows that the persistent speckle is reduced by 5 magnitudes.

## 7. Discussion

We chose here to use a circular eccentric mask in order to break the circular symmetry of the aperture and the speckles. There are many other options for such apertures, especially those designed specifically to reduce stellar speckle [27,28], and they can be optimized for use here provided that they remove the pupil degeneracy, and their light efficiency is high. Such an asymmetric mask for roll deconvolution in space was successfully tested in the laboratory [29]. For the pupil obscuration we have chosen, a side-blocking large circle (Fig. 2), the fractional light loss is almost equal to the fractional narrower pupil dimension. For example, a pupil obstructing 15% of the diameter will cost us about 15% of the light. In terms of throughput and peak throughput [30] this method seems to compare favorably with other co-ronagraphs. We are also looking into hexagonal symmetries in the telescope and ways to remove their pattern.

One of the limitations of planet detection methods is sky background. If the number of photons from the planet is low compared to the sky, and the planet's image is spread over many pixels, the signal to background in a single exposure diminishes the detection probability. If we use only one or a few choice images for each pixel, this is the main limitation. Reducing the resolution might mix back again the persistent speckle into the coarser pixels, and a balance between resolution, step size, and signal must be made.

If a rotating mask is placed at the Lyot aperture of a coronagraph, it is important that it will not scatter stellar light back into the image center. If spatial light modulators can serve for this purpose, they can remove the need for mechanical rotation, and can implement many other options, such as random aperture masks as well as phase masks. Another option is to change the intensity gradually (rather than in a binary mask) in some orientation, by a gradient neutral density filter, and rotate that wedge filter as before. We recently proved that the diffraction pattern has a subtracted weaker component of $J_2^2(r)/r^2$ if the intensity gradient is not too high [31]. Optical vortex coronagraphy [32] is yet another option for less scatter.

If the speckle boiling is slow, the integration time at each orientation step of the camera can be increased accordingly. As we are looking for dim planets, this would reduce the negative bias when looking for the faintest pixels. Saturation of the stellar image is not important, unless the camera suffers from blooming and signal leakage into neighboring pixels. In the opposite case of short exposures and many small rotations, there is no need to accumulate many frames: every new frame only needs to be compared to the stored frame of previous minima in order to update it, then discarded.

As we are employing narrow-band imaging, we can use the spectral Airy ring scaling [25,26] in addition to the pupil rotation. We intend to look into this combination in the future.

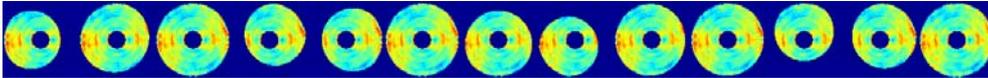

Fig. 9. Changing the occulting aperture size and rotating it at the same time. In this example, the blocked portion grew from 0 up to 24% of the diameter, repeating cyclically 7/3 times the rotation speed.

## 8. Summary

We showed that by using a simple rotating mask we can reach deep nulls in the speckle pattern of a star-planet system, so much so that we can detect faint objects in these nulls very close to the star. Using a characterization of the intensity variations in each image pixel obtained at a few orientations, it is possible to remove most of that stellar signal and remain with the planet signal beneath it. Closer to the image center more orientation steps are required, and the slight residual motion of the stellar diffraction pattern forces us to search for the minima in the intensity, similar to the dark speckle method. Application of a coronagraph will further improve our results.

**Acknowledgements**

ER would like to thank Chris Dainty for his hospitality and his comments, to Nicholas Devaney for helpful discussions, and to Ruth Mackay for help with the lab experiments, all at NUI Galway. Andrei Tokovinin was kind enough to provide us with the wave front errors of the SOAR primary.